\documentclass[aps,prl,twocolumn,amsmath,amssymb,showpacs]{revtex4-1}
\usepackage{graphicx}% Include figure files
\usepackage{dcolumn}% Align table columns on decimal point
\usepackage{bm}% bold math
\usepackage{color}
\newcommand{\hH}{\hat{H}}
\newcommand{\hHso}{\hat{H}_{\rm so}}
\newcommand{\hp}{\hat{p}}
\newcommand{\hx}{\hat{x}}
\newcommand{\hV}{\hat{V}}
\newcommand{\hj}{\hat{j}}
\newcommand{\hU}{\hat{U}}

\newcommand{\comment}[1]{}

\begin{document}

\title{Onsager relations in a two-dimensional electron gas with spin-orbit coupling}

\author{C. Gorini$^a$, R. Raimondi$^b$,  P. Schwab$^{c}$ }
\affiliation{
$^a$Institut de Physique et Chimie des Mat\'eriaux de Strasbourg (UMR 7504),
CNRS and Universit\'e de Strasbourg, 23 rue du Loess, BP 43, F-67034 Strasbourg Cedex 2, France\\
$^b$CNISM and Dipartimento  di Fisica "E. Amaldi", via della Vasca Navale 84, Universit\`a  Roma Tre, 00146 Roma, Italy\\
$^c$Institut f\"ur Physik, Universit\"at Augsburg, 86135 Augsburg, Germany
}

\date{\today}% It is always \today, today,
             %  but any date may be explicitly specified

\begin{abstract}
Theory predicts for the two-dimensional electrons gas with only Rashba spin-orbit interaction 
a vanishing spin Hall conductivity and at the same time
a finite inverse spin Hall effect. We show how these seemingly contradictory results 
are compatible with the Onsager relations: 
the latter do hold for spin and particle (charge) currents in the two-dimensional electron gas, 
although (i) their form depends on the experimental setup and
(ii) a vanishing bulk spin Hall conductivity 
does not necessarily imply a vanishing spin Hall effect. 
We also discuss the situation in which extrinsic spin orbit from impurities is present
and the bulk spin Hall conductivity can be different from zero.
\comment{
Motivated by a recent microscopic calculation [L. Y. Wang et al, arXiv:1111.5082],
We study, in a two-dimensional electron gas with Rashba spin-orbit interaction, 
the generation of spin and charge currents through (1) an inhomogeneous {\it Zeeman} field and
(2) a spin-dependent vector potential in the framework of coupled drift and diffusion equations.
We discuss the general relation between spin and charge currents for the inverse 
spin Hall effect for a two-dimensional electron gas in the presence of Rashba 
spin-orbit interaction by using a $SU(2)$ formulation.
We show that, although the value of the currents may be different depending on the nature of the driving source,  
the relation between the spin and charge current remains universal. 
Furthermore, by allowing also for the extrinsic spin-orbit interaction, 
we find that the charge current response to the stimulus conjugated to the spin current is related to 
the spin current response to the stimulus conjugated to the charge current 
as it is expected on the basis of the Onsager relations.
Our analysis, then, should turn out useful in the interpretation of direct and inverse spin Hall effect experiments.
}
\end{abstract}

% PACS, the Physics and Astronomy
                             % Classification Scheme.
%\keywords{Suggested keywords}%Use showkeys class option if keyword
                              %display desired
\maketitle

It has been repeatedly questioned in the literature whether the Onsager relations \cite{nagaosa2010} 
between direct and inverse spin Hall effect are satisfied\cite{hankiewicz2005,shi2006,jacquod2012}.
In particular it has been argued that with the conventional definition of a spin current --- defined as the product of spin and
velocity operators --- one cannot establish an Onsager relation \cite{shi2006}.
Most recently doubts about their validity have been formulated \cite{wang2012} 
after the prediction of a finite inverse spin Hall effect in the
two-dimensional electron gas with Rashba spin-orbit (SO) coupling \cite{schwab2010}, a system where the spin Hall conductivity 
vanishes \cite{mishchenko2004,inoue2004,rashba2004,raimondi2005,dimitrova2005,chalaev2005}.
In this paper we will cast the SO interaction in terms of non-Abelian $SU(2)$
gauge fields \cite{mathur1992, froehlich1993, tokatly2008, gorini2010, tokatly2010} and show that:
i) Onsager relations \textit{do} hold in the presence of SO coupling, provided 
   the appropriate form of the spin current is used -- crucially, this will depend on the particular measuring
   scheme employed;
ii) a vanishing bulk spin Hall conductivity does \textit{not} imply a vanishing spin Hall or inverse
   spin Hall effect.
We will discuss in some detail the experimental relevance of our results,
which will be shown to be valid in the presence of extrinsic SO coupling from impurities, too.

To begin our discussion, let us imagine a two-dimensional electron gas (2DEG) with SO coupling.
The Hamiltonian is 
\begin{equation} 
\hH = \frac{ \hat{\bf p}^2 }{2m} +  \hHso + V_{\rm imp} (\hat{\bf x}),
\end{equation}
where $V_{\rm imp}$ is a random potential due to impurities, taken to be
$s$-wave scatterers.
Here and throughout a ``hat'' indicates an operator ($\hat{O}$); its corresponding expectation value
will be denoted by the same symbol without ``hat'' ($O$).
For definiteness' sake we choose for $\hHso$ the Rashba SO interaction,
$ \hHso = -\alpha \hp_x \tau^y + \alpha \hp_y \tau^x $, though any other linear-in-momentum
SO term could be handled (see below);  
$\tau^x $ and $\tau^y$ are Pauli matrices, $\alpha$ is the SO coupling constant.
We now add a time dependent perturbation $\hV_1(t)$ of the form
\begin{eqnarray}
 \hV_1(t) & =  & \sum_i  \frac{\hp_i}{m}\left[   e    A_i (t) + \frac{\tau^z}{2}  \eta{ \cal A}_i^z(t) \right]
    \label{eq2} \\
      &=   &   \sum_i  \left[  \hj_i eA_i(t) + \hj_i^z \eta{\cal A}_i^z(t) \right] \label{eq3}
.\end{eqnarray}
The vector potential $A_i(t)$ is related to the electric field via $E_i = - \partial_t A_i$ and is coupled
to the particle current $\hj_i$, whereas ${\cal A}_i^z(t)$ is a fictitious spin dependent vector potential 
which creates a spin electric field
${\cal E}_i^z = - \partial_t {\cal A}_i^z$ and which is coupled to the {\it conventional } spin current $\hj_i^z$;
$\eta$ is a formal $SU(2)$ coupling constant.
Physical mechanisms actually generating this type of spin-dependent vector potential 
are discussed in Ref.~[\onlinecite{wang2012}]. 
Lower [upper] indices indicate real space [SU(2), i.e. spin space] components. 
The Onsager relations connect the spin current generated by an electric field
to the particle current generated by a spin-electric field. 
For the spin Hall effect we conclude from Eq. (\ref{eq3})
\begin{equation} \label{eq4}
j_y(\omega)  = -\sigma^{sH}(\omega) \eta{\cal E}_x^z(\omega)  
\Leftrightarrow j_x^z(\omega) = \sigma^{sH}(\omega) eE_y(\omega)
,\end{equation} 
where $\sigma^{sH}$ is the spin Hall conductivity and $\omega$ is the frequency.
\comment{
Notice that the currents $j_i$ and $j_i^a$ appearing in Eq.(\ref{eq3}) are operators,
while those of Eq.(\ref{eq4}) are expectation values. In general, in the following, operators will only appear in Hamiltonian terms 
and expectation values everywhere else, so no confusion should arise.}
Instead of introducing the electric field ${\bf E}$  
via the vector potential ${\bf A}(t) $, one could, equivalently, choose 
a scalar potential  $\phi(\hat{\bf x}, t ) = -  \mathbf{ E}(t) \cdot \hat{\bf x}  $. 
One could then ask: will the Onsager relations (\ref{eq4}) still hold 
once the spin-electric field is introduced via a spin dependent scalar potential?
With the conventional definition for the spin current introduced in Eq. (\ref{eq3})
the answer is ``yes'' only for vanishing SO coupling. 
This means that for $\alpha = 0$ the time dependent perturbation
\begin{equation}
\label{V2}
 \hV_2(t)  =  - e \phi(\hat{\bf x}, t) - \eta\frac{\tau^z}{2} \Psi^z(\hat{\bf x}, t) ,
\end{equation}
with the spin dependent scalar potential $\Psi^z =-  \mathbf{ \cal E}^z \cdot \hat{\bf x}   $ 
generates the same currents as $\hV_1$.  
Formally, the two cases $\hH_1\equiv \hH+\hV_1,\,\hH_2\equiv \hH+\hV_2$ are connected by a gauge transformation
\begin{equation} \label{eq6}
\alpha=0\;\Rightarrow\; \hH_2 \to \hH_1 = \hU \hH_2 \hU^+ - i \hbar \hU \partial_t \hU^+, 
\end{equation}
with $\hU = \exp[- (i\tau^z/2 \hbar) \eta\chi(\hat{\bf x},t )]$
%\begin{equation} \label{}
% U = \exp\left(- \frac{i\tau^z}{2 \hbar} \eta\chi({\bf x, t} )  \right)
%\end{equation}
and $\partial_t \chi(\hat{\bf x}, t) =  \Psi^z(\hat{\bf x}, t)$.
On the other hand when $\alpha\neq0$ $\hH_1$ and $\hH_2$ are \textit{not} connected by any gauge transformation.  
This can best be seen by writing the SO coupling in terms of a spin-dependent vector potential,
\begin{equation}
\hHso  =  \sum_{i,a}\frac{\hp_i \eta({\cal A}_R)_i^a }{m }  \frac{\tau^a }{2},
\end{equation}
where the subscript $R$ stands for ``Rashba'' and $\eta(\mathcal{A}_R)_x^y=-2m\alpha, \ \eta(\mathcal{A}_R)_y^x=2m\alpha$.
%\begin{equation}
% $\eta(\mathcal{A}_R)_x^y=-2m\alpha, \ \eta(\mathcal{A}_R)_y^x=2m\alpha$
%. \end{equation}
Notice that within this approach a different SO interaction -- e.g. Dresselhaus, a spatially modulated
Rashba and so on -- could be treated just the same and would simply amount to a different choice of $SU(2)$ gauge fields. 
Now the external fields $\hV_1$ and $\hV_2$ are not equivalent any more, since under the gauge transformation (\ref{eq6})
\begin{equation}
\alpha\neq0\;\Rightarrow\;\hH + \hV_2 \to \hH'+\hV_1 \neq \hH + \hV_1,
\end{equation}
i.e. $\hU$ sends $\hV_2\to \hV_1$ and at the same time rotates the background 
Rashba field $\eta{\mathcal{\boldsymbol A}}_R$ sending $\hH \to \hH'$.
Explicitly, to first order in $\chi$ the spin dependent vector potential changes as
\begin{equation}
\label{eq7b}
\eta{\cal A'}^a_i(\hat{\bf x}, t) =\eta{\cal A}^a_i+\eta\hbar^{-1} \chi (\hat{\bf x}, t)  \varepsilon^{abz} {\cal A}^b_i 
-\delta^{az}\eta\nabla_i \chi (\hat{\bf x}, t),
\end{equation}
where $\varepsilon^{abz}$ and $\delta^{az}$ are the fully antisymmetric Ricci tensor and the Kronecker delta.
The Rashba SO term is modified due to the second term on the right-hand side of Eq.~(\ref{eq7b}). 
Physically this is unacceptable: the background Rashba field
has to remain fixed, else we would be describing a different system.  
Such a change can however be absorbed into a redefinition of the spin current:
fixing the background vector potential ${\mathcal{\boldsymbol A}_R}$
requires us to modify the definition of the current coupled to the external perturbation.
To appreciate this point let us take
\begin{equation}
\hV_2(t) =   \frac{\tau^z}{2} \eta{\cal E}_i^z \hx_i,
\end{equation}
and gauge transform $\hH_2$ using $\hU$ previously defined.%Eq.~(\ref{eq7}).
To linear order in the spin-electric field the result is
\begin{equation}
\hH + \hV_2(t) \to \hH +  \hV_1' (t),
\end{equation}
with
\begin{equation}
 \hV'_1(t)  = 
\underbrace{- \frac{i}{\hbar} \left[ \frac{\tau_z}{2} \hx_i , \hH \right]}_{\hat{\tilde{j}}^z_i}
 \underbrace{ \Big( -t\eta {\cal E}_i^z \Big)}_{\eta{\cal A}^z_i}
		\label{defconserved}
,\end{equation}
where $\hat{\tilde{j}}^z_i $ is the {\it conserved} spin current operator suggested in Ref.~[\onlinecite{shi2006}].
Reintroducing the $U(1)$ electric field we can write the equivalent of Eq.~(\ref{eq3})
\begin{equation}
 \hV_1'(t) =   \sum_i  \left[  \hj_i eA_i(t) + \hat{\tilde{j}}_i^z \eta{\cal A}_i^z(t) \right] \label{eq3bis}
\end{equation}
and immediately obtain the Onsager relations
\begin{equation} \label{eq4bis}
j_y(\omega)  = -\tilde{\sigma}^{sH}(\omega) \eta{\cal E}_x^z(\omega)  
\Leftrightarrow \tilde{j}_x^z(\omega) = \tilde{\sigma}^{sH}(\omega) eE_y(\omega).
\end{equation} 
Eqs.~(\ref{eq4}) and (\ref{eq4bis}) are the first main result of this work.
They show that Onsager relations \textit{do} hold in the presence of spin-orbit coupling,
but the quantity reciprocal to the particle current changes depending on the experimental setup
-- i.e. on the way the external spin-electric field is generated.  This means that the
transport coefficient, the spin Hall conductivity, changes too \cite{note1}.

For linear-in-momentum SO interaction the specific form of the spin Hall conductivity
can be computed for any kind of spin electric field
relying on the microscopic formalism developed in Ref.~[\onlinecite{gorini2010}],
which we will now follow. The goal is to verify explicitely the Onsager relations Eqs.~(\ref{eq4}) and (\ref{eq4bis}). 
Let us then focus on the diffusive regime,
in which the equations acquire a remarkable physical transparency. 
Generally, the particle and spin currents are the sum of a diffusion, drift, and a Hall current,
the latter being  responsible for the Hall and  spin Hall effects.
For a system without inversion symmetry, as it is the case for the Rashba model, 
extra terms appear since a homogeneous non-equilibrium spin
density can generate a spin current.  In the $SU(2)$ formulation such extra terms
are automatically built in, and the particle and spin current densities read \cite{gorini2010}
\begin{eqnarray}
{\bf j} &=&  - D \nabla \rho + \sigma {\bf E } 
\label{current1}
-  \frac{\eta \tau}{m} \sum_a {\bf j}^a\times \boldsymbol{\mathcal B}^a,
\\
{\bf j}^a &=& - D [ \tilde{\nabla}s] ^a + \frac{\sigma \eta }{4e} \boldsymbol{\mathcal E}^a - 
\label{current2}
 \frac{\eta \tau}{4m}{\bf j}\times \boldsymbol{\mathcal B}^a
,\end{eqnarray}
when the conventional definition of the spin current is used.  Here, 
$D\equiv v_F^2\tau /2$ is the  diffusion coefficient,
$N_0$ the density of states at the Fermi level, $\tau$ the elastic scattering time 
and $\sigma = -2eN_0 D$, i.e. the electrical
conductivity up to a charge $-e$.  The above equations have been derived under the assumptions of
weak disorder $\epsilon_F \tau \gg \hbar$ and weak SO coupling $\alpha p_F \ll \hbar /\tau$, $\epsilon_F$ and $p_F$
being the Fermi energy and momentum, respectively. In the following, for simplicity, we will use units such that $\hbar =1$.
The $SU(2)$ nature is manifest in the covariant derivative 
$[ \tilde{\nabla}_i  s]^a = \nabla_i s^a - \epsilon^{abc} \eta{\cal A}_i^b s^c$ and in the 
spin dependent electric and magnetic fields 
\begin{eqnarray}
&&
{\mathcal E}_i^a  = -\partial_t {\mathcal A}_i^a - \nabla_i\Psi^a
 - \epsilon^{abc} \eta\Psi^b  {\mathcal A}_i^c ,\label{spin_elec}
\\
&&
{\mathcal B}_i^a = \frac{1}{2}\epsilon_{ijk}\left(
 \nabla_j{\mathcal A}_k^a - \nabla_k{\mathcal A}_j^a 
 -\epsilon^{abc} \eta{\mathcal A}_j^b {\mathcal A}_k^c \right)\label{spin-magn}
.\end{eqnarray}
For the Rashba model there is only one nonvanishing 
field, namely $ \eta{\mathcal B}_z^z = - (2m \alpha)^2$. Adding the external perturbations
$\hV_1$ or $\hV_2$ introduces further fields.
We first consider $\hV_1$, Eq.~(\ref{eq2}), and obtain the additional fields as
$\mathcal{E}_x^z = i\omega \mathcal{A}_x^z,\quad \mathcal{B}^y_z=-(2m\alpha) \mathcal{  A }_x^z$,
%\begin{equation}
%\label{eq14}
%\mathcal{E}_x^z = i\omega \mathcal{A}_x^z,\quad \mathcal{B}^y_z=-(2m\alpha) \mathcal{  A }_x^z,
%\end{equation}
having moved to Fourier space $(\partial_t\to-i\omega,\nabla\to{i\bf q})$ for later convenience.
In linear response to the perturbation $\hV_1$,
the transverse particle current generated by the spin-electric field $\mathcal{E}_x^z$
is (about this  point we disagree with Ref.~[\onlinecite{wang2012}], see also the appendix of this paper)
\begin{equation}
\label{eq16} 
j_y = \frac{\eta \tau}{m}  \mathcal{B}_z^z j_x^z =  4 \gamma j_x^z,
\end{equation}
where the dimensionless number $\gamma = - m \alpha^2 \tau \equiv \gamma_{\rm int}$ 
characterizes the coupling strength between spin and particle currents. 
A non zero spin-charge coupling signals the occurrence of the spin Hall effect \cite{dyakonov2007} 
independently of the spin Hall conductivity being different from zero or not, 
the latter fact depending of the experimental setup and other possible interactions in the Hamiltonian.
The expression for the spin current of Eq. (\ref{current2}) reads
\begin{equation} 
\label{eq18}
j_x^z=-Diq_x s^z +2m\alpha D s^x + \frac{\sigma \eta }{4e}  \mathcal{E}_x^z,
\end{equation}
and in order to find its value we need the spin densities. These can be obtained
by solving the associated diffusion equations, 
which are nothing but the continuity equations for the currents (\ref{current1}) and (\ref{current2}), 
provided the $SU(2)$ covariant derivatives are used\cite{gorini2010}
\begin{equation}
\label{continuity1}
[\tilde{\partial}_t s]^a + [\tilde{\nabla}\cdot{\bf j}] ^a = 0,
\end{equation}
with $[\tilde{\partial}_t s]^a=\partial_t s^a +\epsilon^{abc}\eta\Psi^b s^c$.
In particular the equations 
for the in-plane spin densities in Fourier space are ($\Psi=0$ for the present case of $\hV_1$)
\begin{eqnarray}
\label{continuity2} 
-i\omega s^x + i{\bf q} \cdot {\bf j}^x + 2m \alpha j_x^z &= &0 \\
\label{continuity3} 
-i\omega s^y + i{\bf q} \cdot {\bf j}^y + 2m \alpha j_y^z & = & 0.
\end{eqnarray}
Inserting the Fourier transform of Eqs.~(\ref{current1})-(\ref{current2})
into (\ref{continuity2})-(\ref{continuity3}) 
one obtains in the spatially homogeneous situation 
\begin{equation}
\label{uniform_solution}
j_x^z= \frac{\sigma \eta}{4e} \frac{-i \omega }{- i \omega +\tau^{-1}_{DP}} {\cal E}_x^z,
\end{equation}
where we have introduced the Dyakonov-Perel spin relaxation time $\tau_{DP}^{-1}\equiv(2m\alpha)^2 D$.
Notice that Eq. (\ref{uniform_solution}) is non analytic in $\omega$ and $\tau_{DP}^{-1}$.
In the absence of Rashba SO coupling, i.e. in the limit $\tau_{DP}^{-1}\rightarrow 0$, 
the spin current is given by the spin electric field according to Ohm's law.
When SO coupling is present, the spin current vanishes in the DC limit, i.e. $\omega \rightarrow 0$. 
In the appendix this is shown explicitly by evaluating the
Kubo formula diagrammatically.
Relation (\ref{eq16}) yields the particle-current response to the 
spin-electric field and,
to leading order in ${\cal B}_z^z$,
\begin{equation}
\label{spin_hall}
\sigma^{sH}(\omega ) = -\frac{\gamma \sigma }{e}   \frac{-i \omega}{-i\omega +\tau^{-1}_{DP}}
.\end{equation}
As required by the Onsager relations (\ref{eq4}) this  agrees with the spin Hall conductivity determined by the  
response of the conventionally defined spin current to the electric field.
The latter result can be obtained by combining the expression for
the spin current (\ref{current2}) with the continuity equation (\ref{continuity2}).

We can now follow the same route while considering the external perturbation $\hV_2$, Eq.~(\ref{V2}),
with an $x$-dependent spin-scalar potential $\Psi^z(\hx,t)$.
The latter introduces the following fields ${\cal E}_x^x={\cal E}_y^y=-2m\alpha\Psi^z,\;{\cal E}_x^z=-iq_x\Psi^z$. 
%\begin{equation}
%{\cal E}_x^x={\cal E}_y^y=-2m\alpha\Psi^z,\;{\cal E}_x^z=-iq_x\Psi^z.
%\end{equation}
Our system is now homogeneous only along $y$, and the diffusion equations read
\begin{eqnarray}
-i\omega s^x & = &  (-Dq_x^2 -\tau_{DP}^{-1})s^x + 4m\alpha D  iq_x [ s^z - (N_0/2)\eta\Psi^z ],\nonumber \\
-i\omega s^y & = &  (-Dq_x^2 -\tau_{DP}^{-1})s^y, \label{eq_dif1}\\
-i\omega s^z & = &  (-Dq_x^2 -2\tau_{DP}^{-1})[s^z-(N_0/2)\eta\Psi^z]-4m\alpha D iq_x s^x,\nonumber
\end{eqnarray}
where 
we have ignored all terms that are quadratic in the external field $\Psi^z$; 
notice that in the absence of $\Psi^z$ no spin polarization exists, thus the spin density is itself
at least ${\cal O}(\Psi^z)$.
Solving Eqs.~(\ref{eq_dif1}) for a homogeneous but frequency dependent spin-electric field we find
\begin{equation}
j_x^z=\frac{\sigma \eta }{4e} \frac{-i \omega }{- i \omega +2\tau^{-1}_{DP}}
\frac{-i\omega-\tau^{-1}_{DP}}{-i\omega +\tau^{-1}_{DP}}{\cal E}_x^z,
\end{equation}
and with Eq.~(\ref{eq16}) we conclude that the spin Hall conductivity is 
\begin{equation}
\label{sigmatilde}
\tilde{\sigma}^{sH}(\omega ) =-\frac{\gamma \sigma}{e}
\frac{-i \omega }{- i \omega +2\tau^{-1}_{DP}}\frac{-i\omega-\tau^{-1}_{DP}}{-i\omega +\tau^{-1}_{DP}}.
\end{equation}
According to the Onsager relations (\ref{eq4bis}),
the reciprocal quantity to the inverse spin Hall current $j_y$
is the conserved spin current $\tilde{j}_x^z$ generated by an homogeneous and
frequency dependent electric field along $y$,
\begin{equation}
\label{conserved}
\tilde{j}_x^z = \lim_{q_x\rightarrow0} \frac{\omega}{q_x}s^z.
\end{equation}
The above relation follows from the continuity equation for the conserved current and from the observation that only the longitudinal current is needed for the Hall response.
The diffusion equations to solve are now [we drop terms ${\cal O}(q_x^2)$]
\begin{eqnarray}
-i\omega s^x & = &  -\tau_{DP}^{-1}s^x+2m\alpha\left[2Diq_x s^z+\gamma\sigma E_y\right] \nonumber\\
-i\omega s^y & = &  -\tau_{DP}^{-1}s^y \label{diffusion}\\
-i\omega s^z & = &  -2\tau_{DP}^{-1}s^z-iq_x\left[2(2m\alpha)D s^x-\gamma\sigma E_y\right]\nonumber.
\end{eqnarray}
Their solution yields $\tilde{j}_x^z=\tilde{\sigma}^{sH}(\omega)eE_y$
with $\tilde{\sigma}^{sH}(\omega)$ given by (\ref{sigmatilde}), thus
verifying the validity of Eq.~(\ref{eq4bis}).

It is now worthwhile investigating the robustness of the above results 
to the presence of extrinsic SO interaction arising from impurities,
since the latter are usually present in real samples
and in this case the static spin Hall conductivity $\sigma^{sH}(\omega\to0)$ 
is different from zero\cite{raimondi2009,raimondi2012}.
To this end we add to the Hamiltonian the extrinsic term
\begin{equation}
\label{ext_so}
\hH_{\rm extr}=-\frac{\lambda_0^2}{4} {\boldsymbol \tau} \times \nabla V_{\rm  imp} (\hat{\bf x}) \cdot \hat{\bf p},
\end{equation}
where ${\boldsymbol \tau}$ is the vector of Pauli matrices and
$\lambda_0$ is the effective Compton wavelength describing the SO coupling in the system. 
The extrinsic SO interaction (\ref{ext_so}) modifies the theory only  in two main aspects.
First, the presence of the extrinsic SO scattering introduces the Elliott-Yafet spin relaxation time, 
$\tau_s$,  so that Eq. (\ref{continuity2}) is modified to
\begin{equation}
\label{eq27}
-i\omega s^x +i{\bf q}\cdot{\bf j}^x+2m\alpha j_x^z =-\tau^{-1}_s s^x,
\end{equation}
with $\tau_s =\tau (\lambda_0 p_F/2)^{-4}$. 
The second ingredient is that the parameter $\gamma$ entering Eq. (\ref{eq16}) 
acquires a contribution from the skew-scattering and side-jump mechanisms $\gamma =\gamma_{\rm  int}+\gamma_{ss}+ \gamma_{sj}$
%\begin{equation}
%\label{eq28}
%\gamma =\gamma_{\rm  int}+\gamma_{ss}+ \gamma_{sj},
%\end{equation}
where $\gamma_{\rm int}=-m \alpha^2\tau$ as before, while $\gamma_{sj}=(\lambda_0/2)^2 (m/\tau)$ and 
$\gamma_{ss}=-(\lambda_0 p_F/4)^2 (2\pi N_0 v_0)$, $v_0$ being the impurity scattering amplitude, 
see Ref.~[\onlinecite{raimondi2010}] for technical details. 
One can now proceed as before and check that in linear response to $\hV_1$ and $\hV_2$
the relations (\ref{eq4}) and (\ref{eq4bis}) still hold, with the spin Hall conductivities
\begin{eqnarray}
\label{sigmatotal}
\sigma^{sH}(\omega) &=& - \frac{\gamma \sigma}{e} \frac{-i \omega + \tau^{-1}_s }{-i\omega + \tau^{-1}_{DP}+\tau^{-1}_s}
\\
\label{sigmatildetotal}
\tilde{\sigma}^{sH}(\omega ) &=& -\frac{\gamma \sigma}{e}
\frac{-i \omega }{- i \omega +2\tau^{-1}_{DP}}
\frac{-i\omega-\tau^{-1}_{DP}+\tau^{-1}_s}{-i\omega +\tau^{-1}_{DP}+\tau^{-1}_s}.
\end{eqnarray}
We wish to stress two important points.  
First, in obtaining the above we could still exploit Eq.~(\ref{conserved}),
since the spin current $\tilde{\bf j}^a$ introduced in Eq.~(\ref{defconserved})
is by definition conserved with respect to the full background field $\hHso+\hH_{\rm extr}$\cite{note3}.
Second, and experimentally important, 
in the absence of intrinsic SO coupling,
one has to take the $\alpha \rightarrow 0$ limit first, so that   $\sigma^{sH}(\omega) = \tilde{\sigma}^{sH} (\omega)$,
i.e. the two experimental setups corresponsing to $\hV_1$ and $\hV_2$ become equivalent, 
since the out-of-plane spin density becomes a conserved quantity\cite{note2}.
This is not the case in the presence of both intrinsic and extrinsic SO mechanisms,
since $\hV_1$ is capable of sustaining a steady state bulk spin Hall current, whereas $\hV_2$ is not. 
It must be pointed out that by using the formula for the conserved spin current derived by
Sugimoto et al.\cite{sugimoto2006} (cf. their Eq.(9))  with the self-energy inclusive of the spin-orbit from impurities
(cf. Eq.(\ref{ext_so}) and Ref.[\onlinecite{raimondi2010}] for details), one finds a zero spin Hall conductivity
in agreement with the zero-frequency limit of Eq.(\ref{sigmatildetotal})\cite{note4}.

The relevance of our results with respect to available experiments 
is worth a more detailed discussion. 
Theory tells that in the pure Rashba case the bulk spin Hall conductivity vanishes,
it is neither possible to drive a spin current by a uniform and weakly time dependent 
electric field, nor to drive a charge current 
by (i) a uniform but weakly time dependent spin-vector potential
(ii) a weakly space dependent but static spin-scalar potential.
On the other hand when both intrinsic and extrinsic SO interaction are present, 
the bulk spin Hall conductivity can be different from zero.
To distinguish which spin current is excited in a given setup, according to Eqs.(\ref{sigmatotal}-\ref{sigmatildetotal}), 
one should perform an inverse spin Hall effect experiment and measure the frequency dependent induced voltage. 
Alternatively, one could consider a purely electrical measurement looking at the frequency-dependent non local resistance 
in a four-probe set up as that considered in Ref. [\onlinecite{brune2010}].
A linear frequency behavior signals the excitement of the conserved current. 
A cubic Dresselhaus term has a similar effect\cite{wang2012,malshukov2005}.
However even a vanishing bulk spin Hall conductivity does not imply 
the absence of the spin Hall effect and its inverse. 
The spin Hall effect and an induced edge spin-polarization are present 
close to a interface where non-spin-polarized carriers 
are injected into the Rashba 2DEG. This has been predicted first in Ref. [\onlinecite{mishchenko2004}]
and verified numerically in Ref. [\onlinecite{raimondi2006}].
This is also manifest in the expression for the spin current, Eq.~(\ref{current2}),
since when spin polarization is negligibly small the current becomes
\begin{equation}
{\bf j}^a =-  \frac{\eta \tau }{4m} {\bf j } \times \boldsymbol{\mathcal B}^a
.\end{equation}
For the inverse spin Hall effect the situation is analogous. 
In an experiment such as the one of Ref.~[\onlinecite{wunderlich2009}]
no spin-electric field is applied to the samples.
Instead a circularly polarized laser beam is used to create electron-hole pairs at a {\it p-n} 
junction between a 2DEG and a two-dimensional hole gas.  
With the junction suitably biased, spin polarized electrons are injected in the 2DEG,
so the spin current ${\bf j}^a$ at the interface is directly determined by the experimental setup and thus
creates a Hall signal,
\begin{equation}
{\bf j }_{\rm Hall}  = - \frac{\eta \tau}{m} \sum_a {\bf j}^a \times \boldsymbol{\mathcal B}^a.
\end{equation}

To conclude, we have shown the existence of 
Onsager relations connecting electric to spin-electric stimuli in a two-dimensional
electron gas with spin-orbit coupling.
In order to be explicit we focused on the Rashba model, but the non-Abelian formulation
employed can be used for any linear-in-momentum SO interaction,
possibly slowly varying in time and space, too.  
%Within such an approach the non-conservation of spin
%is automatically built into the equations and never leads to any ambiguity
%-- in particular concerning dissipation.
Quite important from the experimental point of view, 
the Onsager relations obtained are robust to the inclusion of extrinsic SO coupling from impurities
and their specific form depends crucially on the measuring scheme employed.

We acknowledge financial support from the EU through Grant. No. PITN-GA-2009-234970, 
from the Deutsche Forschungsgemeinschaft through TRR80 and SPP 1285 and
from the French National Research Agency ANR, Project No. ANR-08-BLAN-0030-02.

%\section{Appendix}
\textit{APPENDIX} - In the main text we have shown that a static spin-electric field introduced via a perturbation  $\hV_1$ does not create a spin current,
a result which does  not agree with Eq. (15) of Ref. [\onlinecite{wang2012}].
To further support this statement we show here how to obtain this result with a different method, 
namely by evaluating the suitable Kubo formula for the {\it spin current-spin current} 
correlation function. By using the notation of Ref. \onlinecite{raimondi2005} we have that 
\begin{equation}
j_x^z=\sigma_{xx}^{zz}{\cal E}^z_x, \  \    \sigma_{xx}^{zz}=-\frac{1}{2\pi}\sum_{\bf p} {\rm Tr}  \left[
G^A \hat{J}_x^z G^R \hat{j}_x^z\right],
\end{equation}
where $\hat{j}_x^z$ and $\hat{J}_x^z$ are the bare and dressed  spin current vertices $\hat{j}_x^z =(\hat{p}_x/2m)\tau^z$, $ \hat{J}_x^z=\hat{j}_x^z +\hat{\Gamma}_x^z$.
%\begin{equation}
%j_x^z =(p_x/2m)\tau^z,    \  J_x^z=j_x^z +\Gamma_x^z.
%\end{equation}
We then obtain
\begin{equation}
\sigma_{xx}^{zz}=-\frac{\sigma}{4e^2}\frac{1- (2\alpha p_F \tau /v_F){\rm Tr}(\tau^x \hat{\Gamma}_x^z)}{1+(2\alpha p_F \tau)^2}.
\end{equation}
The vertex corrections to the spin current vertex have been evaluated in Ref.~\cite{wang2006} 
with the result $\hat{\Gamma}_x^z =v_F (4\alpha p_F \tau)^{-1}\tau^x$. 
One then obtains the vanishing of the spin current.


\begin{thebibliography}{999}
\bibitem{note1} Notice that if $\hH_1$ and $\hH_2$ were indeed connected by a $SU(2)$ gauge transformation
the inverse spin Hall current would trivially be the same in both cases,
since the particle current is a $SU(2)$ invariant.
\bibitem{note2} Technically, this can be traced back to the fact that $H_{\rm extr}$ does not change
the covariant derivatives, see Ref.~[\onlinecite{raimondi2012}].
\end{thebibliography}
\end{document}